\documentclass[fleqn,usenatbib]{mnras}

\usepackage{newtxtext,newtxmath}

\usepackage[T1]{fontenc}

\DeclareRobustCommand{\VAN}[3]{#2}
\let\VANthebibliography\thebibliography
\def\thebibliography{\DeclareRobustCommand{\VAN}[3]{##3}\VANthebibliography}

\usepackage{graphicx}	
\usepackage{amsmath}	
\usepackage{textcomp}	
\usepackage[dvipsnames]{xcolor}
\usepackage{tikz}
\usetikzlibrary{arrows.meta,
				 shapes,
                chains,
                positioning,
                shapes.geometric}
\tikzstyle{arrow}=[draw, -latex]

\newcommand{\appropto}{\mathrel{\vcenter{
  \offinterlineskip\halign{\hfil$##$\cr
    \propto\cr\noalign{\kern2pt}\sim\cr\noalign{\kern-2pt}}}}}

\newcommand{\Mp}{M_\mathrm{p}}
\newcommand{\Mth}{M_\mathrm{th}}
\newcommand{\hp}{h_\mathrm{p}}
\newcommand{\Hp}{H_\mathrm{p}}
\newcommand{\Rp}{R_\mathrm{p}}
\newcommand{\Pp}{P_\mathrm{p}}
\newcommand{\de}{\mathrm{d}}

\newcommand{\const}{\mathrm{const.}}

\newcommand{\tvrt}{\tau_\mathrm{vrt}}
\newcommand{\tp}{\tau_\mathrm{p}}

\newcommand\ba{\begin{eqnarray}}
\newcommand\ea{\end{eqnarray}}

\defcitealias{Goodman2001}{GR01}
\defcitealias{Rafikov2002}{Rafikov 2002}
\defcitealias{Cimerman2022}{CR23}

\title[Vortex weighing and dating]{Vortex weighing and dating of planets in protoplanetary discs}

\author[R. R. Rafikov and N. P. Cimerman]{
Roman R. Rafikov$^{1,2}$\thanks{E-mail: rrr@damtp.cam.ac.uk (RRR)}, Nicolas P. Cimerman$^{1}$
\\
$^{1}$Department of Applied Mathematics and Theoretical Physics, University of Cambridge, Wilberforce Road, Cambridge CB3 0WA, UK\\
$^{2}$Institute for Advanced Study, Einstein Drive, Princeton, NJ 08540, USA
}

\date{Accepted XXX. Received YYY; in original form ZZZ}

\pubyear{2022}

\begin{document}
\label{firstpage}
\pagerange{\pageref{firstpage}--\pageref{lastpage}}
\maketitle

\begin{abstract}
High-resolution sub-mm observations of some protoplanetary discs reveal non-asixymmetric features, which can often be interpreted as dust concentrations in vortices that form at the edges of gaps carved out by the embedded planets. We use recent results on the timescale for the planet-driven vortex development in low-viscosity discs to set constraints on the mass and age of a planet producing the vortex. Knowledge of the age of the central star in a vortex-bearing protoplanetary disc system allows one to set a lower limit on the planetary mass at the level of several tens of $M_\oplus$. Also, an independent upper limit on the planetary mass would constrain the planetary age, although given the current direct imaging detection limits this constraint is not yet very stringent (it is also sensitively dependent on the disc scale height). These results can be extended to account for the history of planetary mass accretion if it is known. We apply our calculations to several protoplanetary discs harbouring vortex-like features as revealed by ALMA and set limits of $(30-50)M_\oplus$ (for disc aspect ratio of $0.1$) on the minimum masses of putative planets that could be responsible for these vortices. Our vortex-based method provides an independent way of constraining the properties of embedded planets, complementary to other approaches.
\end{abstract}

\begin{keywords}
hydrodynamics -- instabilities -- shock waves -- accretion discs -- planets and satellites: formation -- methods: numerical
\end{keywords}




\section{Introduction}
\label{sec:intro}


Observations of protoplanetary discs (PPDs) in dust continuum emission with ALMA revealed a variety of substructures \citep{Andrews2020}, including axisymmetric gaps and rings as well as non-axisymmetric clumps and arcs.  An intriguing possibility is that these features could be produced by young embedded planets. In particular, gravitational coupling between a massive planet and the disc is known to result in formation of observable gaps around the planetary orbit \citep{Pap1984,Rafikov2002II}.  The evolution of vortensity (potential vorticity) at the edges of these gaps \citep{Lin2010,Dong2011II,Cimerman2021} due to shock dissipation of the planet-driven density waves \citep{Goodman2001,Rafikov2002} can trigger the Rossby Wave Instability \citep[RWI,][]{Lovelace1999} resulting in the formation of fluid vortices at these locations. Dust accumulation inside the vortices \citep{Barge1995} naturally leads to  observable non-axisymmetric arcs and lobes.

Formation of these structures does not necessarily require massive (Jovian) planets. For example, it was shown \citep{Dong2017,Bae2017,Miranda2019II,Miranda2020I,Miranda2020II} that  multiple visible gaps and rings in the dust distribution can result from nonlinear damping of multiple spirals triggered by a single sub-Jovian mass planet in a low viscosity disc. The mass of the planet $M_\mathrm{p}$ can in fact be below the so-called \textit{thermal mass} defined as $\Mth=\left(\Hp/\Rp\right)^3 M_\star=\hp^3 \, M_\star$, where $\Hp$ is the disc scale height at the planetary distance $\Rp$, $\hp=\Hp/\Rp$ is the disc aspect ratio there, and $M_\star$ is the stellar mass. Emergence of vortices at the edges of planetary gaps also does not require massive planets if the disc is almost inviscid \citep{Hammer2021,Hallam2020},  and sub-$\Mth$ mass planets can easily trigger them \citep[][hereafter \citetalias{Cimerman2022}]{Cimerman2022}.

In this study we focus on non-axisymmetric disc features which can be interpreted as planet-induced vortices (other ways to produce vortices are mentioned in Section \ref{sec:obs}). Their development is not an instantaneous process as the evolution of vortensity near the planetary orbit towards the RWI takes a certain amount of time. As we show in this work,  this fact can be exploited to set useful constraints on the mass and/or age of a putative planet responsible for production of the observed vortices in a PPD. This method relies on the recent calculation of \citetalias{Cimerman2022} who studied the development of vortices in inviscid PPDs. In that work, we showed that the time it takes for vortices to emerge at the edge of the gap carved out by a sub-$\Mth$ mass planet can be approximated as
\begin{align}
\tvrt & \approx A\, \Pp \left( \frac{\Mp}{\Mth} \right)^\alpha \hp^\beta,~~~~\mbox{where}
\label{eq:taufit}\\
A & \approx 1.6, ~~~\alpha\approx -2.7,~~~\beta\approx -0.86,
\label{eq:pars}
\end{align}
and $\Pp$ is the orbital period at $\Rp$.  This result assumes $\Mp$ to be fixed in time.  Interestingly, \citetalias{Cimerman2022} found $\tvrt$ to only weakly depend on the radial profile of surface density in the disc.

We describe the general idea of our method in Section \ref{sec:obs} and show how it can constrain the mass and age of the planet in Section \ref{sec:weighting} and Section \ref{sec:dating}, respectively. In Section \ref{sec:growth} we extend our constraints to the case of a planet accreting its mass over an extended period of time. We apply our results to observed PPDs in Section \ref{sec:real} and discuss them in Section \ref{sec:disc}.


\section{General idea of the method}
\label{sec:obs}


Let us suppose that observations of dust continuum emission reveal a gap in a PPD, together with a non-axisymmetric lobe or arc at the gap edge, indicative of a vortex which traps dust grains at this location \citep[e.g.][]{vdM2016,Kraus2017,Dong2018,Perez2018}. We will interpret this observation by assuming that a planet (not necessarily directly visible) is located within the gap and is responsible for creating both the gap and the vortex (via the RWI). The spatial association of a vortex with an adjacent gap provides strong support to this interpretation and makes some other possibilities for triggering vortices, e.g. global baroclinic instability \citep{Klahr2003}, convective overstability \citep{Teed2021}, vertical shear instability \citep{Richard2016} less attractive.

We assume the disc viscosity to be low (essentially inviscid), consistent with many observations of PPDs \citep{Pinte2016,Rafikov2017,Flaherty2020}. For now we will also assume that $\Mp$ has been constant ever since the planet appeared in the disc, a constraint that we will relax in  Section \ref{sec:growth}. With these conditions fulfilled, the result (\ref{eq:taufit})-(\ref{eq:pars}) applies.  We can then use the observation of the vortex to set a constraint on a particular combination of the planetary mass $\Mp$ and the planetary age $\tp$ --- the time that has passed since the planet has reached its final mass $\Mp$. 

\begin{figure}
\centering
	\includegraphics[width=0.49\textwidth]{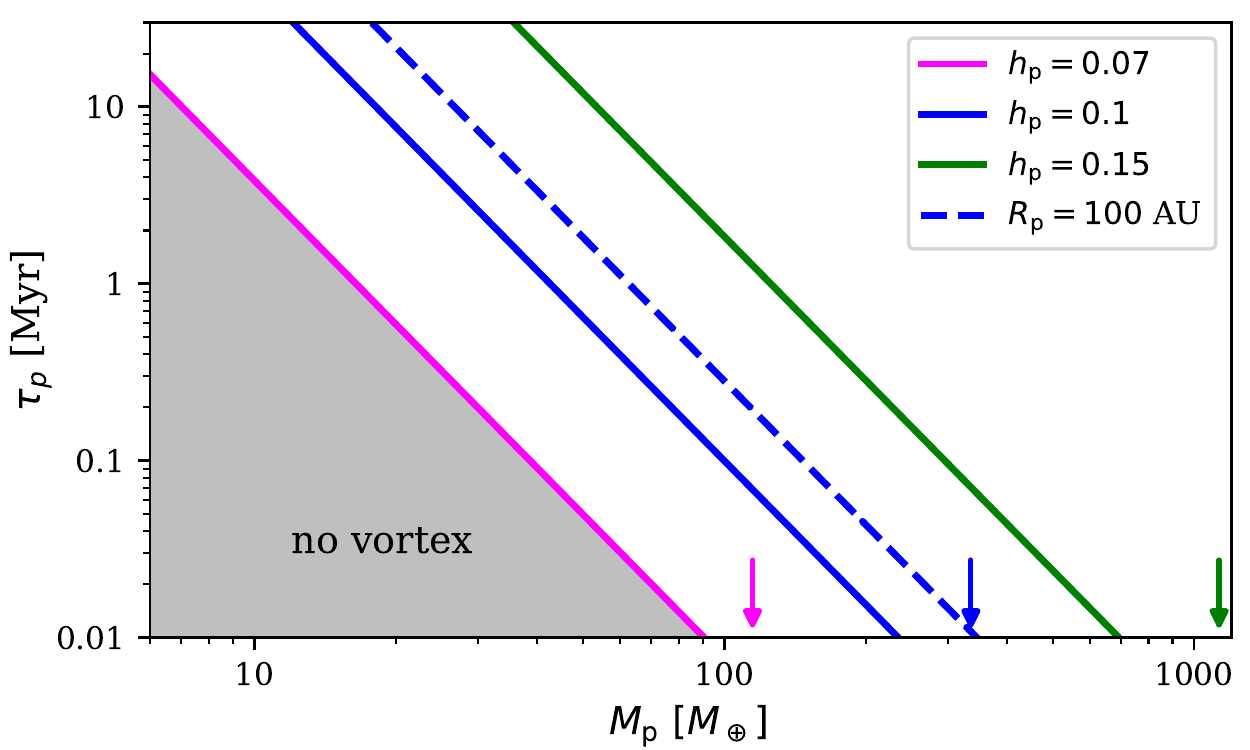}
\vspace*{-0.5cm}
\caption{Combined constraint (\ref{eq:combined_val}) on the planetary mass $\Mp$ and age $\tau_\mathrm{p}$, shown for different parameters of a system with $M_\star=M_\odot$. Solid lines are for $\Rp=50$ AU and $\hp=0.07$ (fuchsia), $\hp=0.1$ (blue), $\hp=0.15$ (green). Blue dashed line is for $\Rp=100$ AU, $\hp=0.1$. Grey shaded region is excluded as no vortices should appear in this part of the parameter space (bounded by $\hp=0.07$ curve for illustration). Arrows indicate $\Mth$ calculated using $\hp$ corresponding to the arrow color (same as in the legend). Constraint (\ref{eq:combined_val}) --- solid curves --- is strictly valid only for $\Mp\lesssim \Mth$.
}
\label{fig:age}
\end{figure}

Indeed, the observation of a vortex at the gap edge implies that the RWI had enough time to fully develop into the non-linear stage in that region, i.e. that 
\begin{align}
\tau_\mathrm{p}>\tvrt.
\label{eq:times}
\end{align}
Together with equation (\ref{eq:taufit}) this leads to the following combined constraint on $\tp$ and $\Mp$: 
\begin{align}
\tp\Mp^{-\alpha}> A \Pp \Mth^{-\alpha} \hp^\beta.
\label{eq:combined}
\end{align}
With fit parameters (\ref{eq:pars}) we can write this in physical units as
\begin{align}
\frac{\tau_\mathrm{p}}{\rm Myr}\left(\frac{\Mp}{10^2M_\oplus}\right)^{2.7}> 0.11 \left(\frac{\Rp}{50\mathrm{AU}}\right)^{1.5}\left(\frac{M_\star}{M_\odot}\right)^{2.2}\left(\frac{\hp}{0.1}\right)^{7.2}. 
\label{eq:combined_val}
\end{align}
This condition must be fulfilled whenever a vortex is observed at the gap edge.  It is illustrated in Fig. \ref{fig:age} for several values of $\hp$ and $\Rp$.

In a similar vein, the {\it absence} of vortex-like structures at the edges of a visible gap in a disc might be interpreted as meaning that $\tp<\tvrt$, i.e. that planet-driven accumulation of vortensity has not yet led to RWI. If that were the case, the inequality in the constraint (\ref{eq:combined})-(\ref{eq:combined_val}) would change its sign. However, this possibility has an important caveat as the absence of a vortex may also be interpreted differently: it could have formed at the gap edge earlier but then got destroyed through one of the processes that tend to destabilize vortices once they evolve into the nonlinear regime: the elliptical instability \citep{Lesur2009}, baroclinic effects \citep{Rometsch2021,FungOno2021}, dust feedback \citep{Fu2014}, etc. Also, vortex formation may have been delayed or suppressed altogether if disc viscosity is sufficiently high \citep{Hammer2017,Hallam2020}. Thus, the lack of a vortex near a planetary gap cannot be unambiguously interpreted as meaning that the embedded planet did not get a chance to create it, i.e. that $\tp<\tvrt$. For that reason (and unlike \citealt{Hallam2020}) in the following we will not draw any conclusions from the {\it absence} of vortices at the edges of putative planetary gaps found in sub-mm observations.

We will now show how equations (\ref{eq:combined}) \& (\ref{eq:combined_val}) can be used to separately constrain  $\Mp$ or $\tau_\mathrm{p}$.


\section{Vortex weighing of planets}
\label{sec:weighting}

\begin{figure}
\centering
	\includegraphics[width=0.49\textwidth]{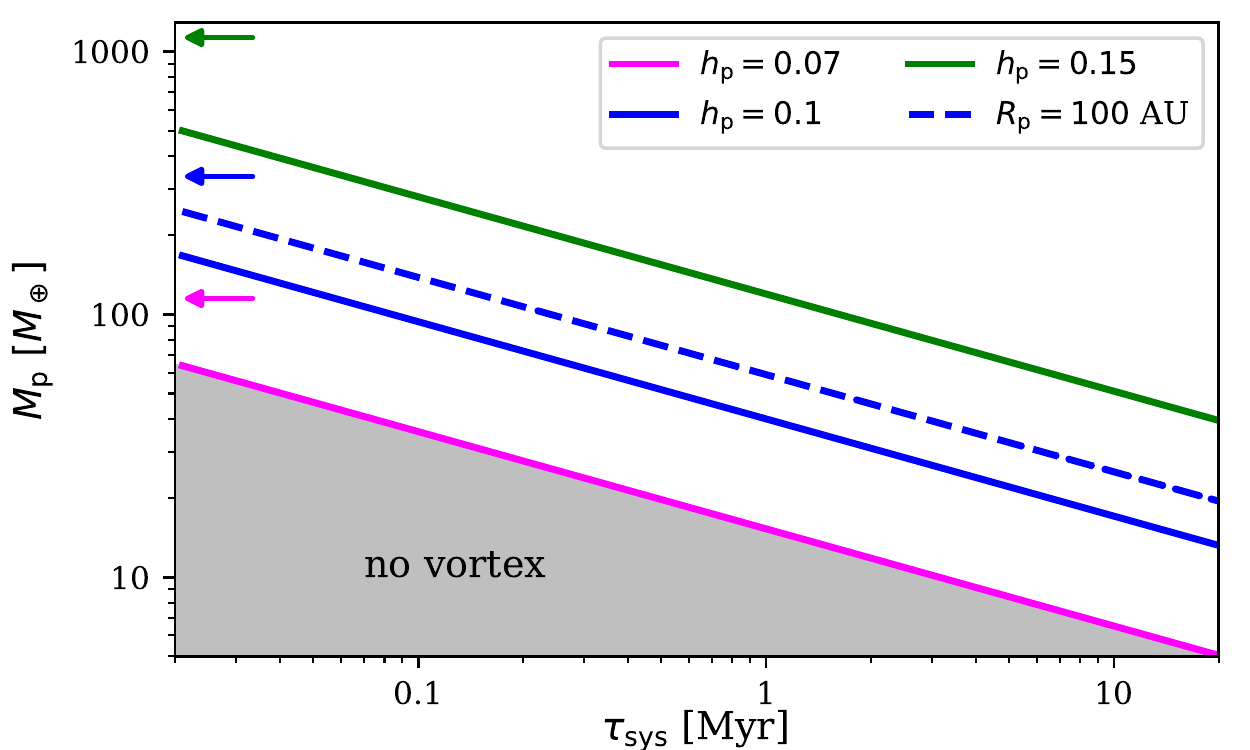}
\vspace*{-0.5cm}
\caption{Mass constraint (\ref{eq:M-constr}), (\ref{eq:M-vrt}) as a function of the system (stellar) age $\tau_\mathrm{sys}$. Meaning of curves, arrows and shading are the same as in Fig. \ref{fig:age}.
}
\label{fig:mass}
\end{figure}

Let us suppose that the age (time since formation) of the protostar-disc system $\tau_\mathrm{sys}$ is known, e.g. from isochrone fitting of the characteristics of the central star.
This is usually the case at some level of accuracy. Since, obviously, the planet is younger than its parent star, one must have $\tau_\mathrm{sys}>\tp$. However, the presence of a vortex at the gap edge means that the inequality (\ref{eq:times}) is also fulfilled, which necessarily implies that $\tau_\mathrm{sys}>\tvrt$. Using equation (\ref{eq:combined}), this condition can be converted into a {\it lower limit} on $\Mp$:
\begin{align}
    \Mp>M_\mathrm{vrt}=\Mth\left(A\,\hp^\beta\frac{\Pp}{\tau_\mathrm{sys}}\right)^{-1/\alpha}.
	\label{eq:M-constr}
\end{align}
In physical units,
\begin{align}
    M_\mathrm{vrt}\approx 40M_\oplus
    \left(\frac{\tau_\mathrm{sys}}{\mbox{Myr}}\right)^{-0.37}
    \left(\frac{\Rp}{50\mbox{AU}}\right)^{0.56}
    \left(\frac{\hp}{0.1}\right)^{2.7}
    \left(\frac{M_\star}{M_\odot}\right)^{0.81}.
	\label{eq:M-vrt}
\end{align}
Note a strong dependence of $M_\mathrm{vrt}$ on $\hp$, but a rather weak scaling with $\tau_\mathrm{sys}$. This constraint is illustrated in Fig. \ref{fig:mass}. 

Note that for the mass constraint (\ref{eq:M-constr})-(\ref{eq:M-vrt}) to be valid, the timescale fit (\ref{eq:taufit}) should be justified in the first place. For this to be the case, the planetary mass must be in the sub-thermal mass regime. One can easily show that
\begin{align}
    \frac{M_\mathrm{vrt}}{\Mth}\approx 0.13
    \left(\frac{\tau_\mathrm{sys}}{\mbox{Myr}}\right)^{-0.37}
    \left(\frac{\Rp}{50\mbox{AU}}\right)^{0.56}
    \left(\frac{\hp}{0.1}\right)^{-0.32}
    \left(\frac{M_\star}{M_\odot}\right)^{-0.19},
	\label{eq:M-vrt-th}
\end{align}
i.e. the condition $\Mp\lesssim\Mth$ should be not difficult to satisfy in general (in Figs. \ref{fig:age},\ref{fig:mass} we illustrate the values of $\Mth$ with arrows). Thus, we expect $M_\mathrm{vrt}$ to provide a lower limit on $\Mp$ quite generally. 

The constraint (\ref{eq:M-constr})-(\ref{eq:M-vrt}) can be improved (i.e. $M_\mathrm{vrt}$ increased) if we had some independent way to set an upper limit on $\tp$, which is lower than the system age $\tau_\mathrm{sys}$. In practice, however, such refined information on $\tp$ may be difficult to obtain.


\section{Vortex dating of planets}
\label{sec:dating}


One can also turn the argument around and assume that, in addition to observing a vortex adjacent to a gap, we also know the mass $\Mp$ of the gap-opening planet --- either via atmospheric modelling if the planet is visible, or through indirect dynamical measurements if it has not been imaged. We can then use the presence of the vortex to set a {\it lower} limit on the planetary age $\tau_\mathrm{p}$ via equation (\ref{eq:times}), in which 
\begin{align}
   \tau_\mathrm{vrt}\approx 10^5\mbox{yr}\left(\frac{\Mp}{10^2M_\oplus}\right)^{-2.7}
    \left(\frac{\Rp}{50\mbox{AU}}\right)^{1.5}
    \left(\frac{\hp}{0.1}\right)^{7.2}
    \left(\frac{M_\star}{M_\odot}\right)^{2.2}.
	\label{eq:T_vrt}
\end{align}
If only an upper limit $M_\downarrow$ on planetary mass is available to us, $\Mp<M_\downarrow$ (e.g. from non-detection of the planet through near-IR imaging), then one should use $M_\downarrow$ instead of $\Mp$ in (\ref{eq:T_vrt}). Solid and dashed lines in Fig. \ref{fig:age} give $\tau_\mathrm{vrt}$ (as a function of $\Mp$ or $M_\downarrow$) for different values of $\hp$ and $\Rp$.

A constraint on $\tau_\mathrm{p}$ would be extremely useful for understanding the timing of planet formation. It can also serve as a consistency check for calculations of planetary evolution post-formation, since the present day temperature and luminosity of the planet are themselves functions of its age $\tau_\mathrm{p}$ \citep[e.g.][]{Linder2019}, see Section \ref{sec:disc}. Unfortunately, the accuracy of the lower limit (\ref{eq:times}) \& (\ref{eq:T_vrt}) may be somewhat compromised by the uncertainties in the determination of various parameters that enter it, e.g. $\Mp$ and, especially, $\hp$, given how steeply $\tau_\mathrm{vrt}$ scales with them. 


\section{Accounting for accretion history of a planet}
\label{sec:growth}

\begin{figure*}
\centering
	\includegraphics[width=0.9\textwidth]{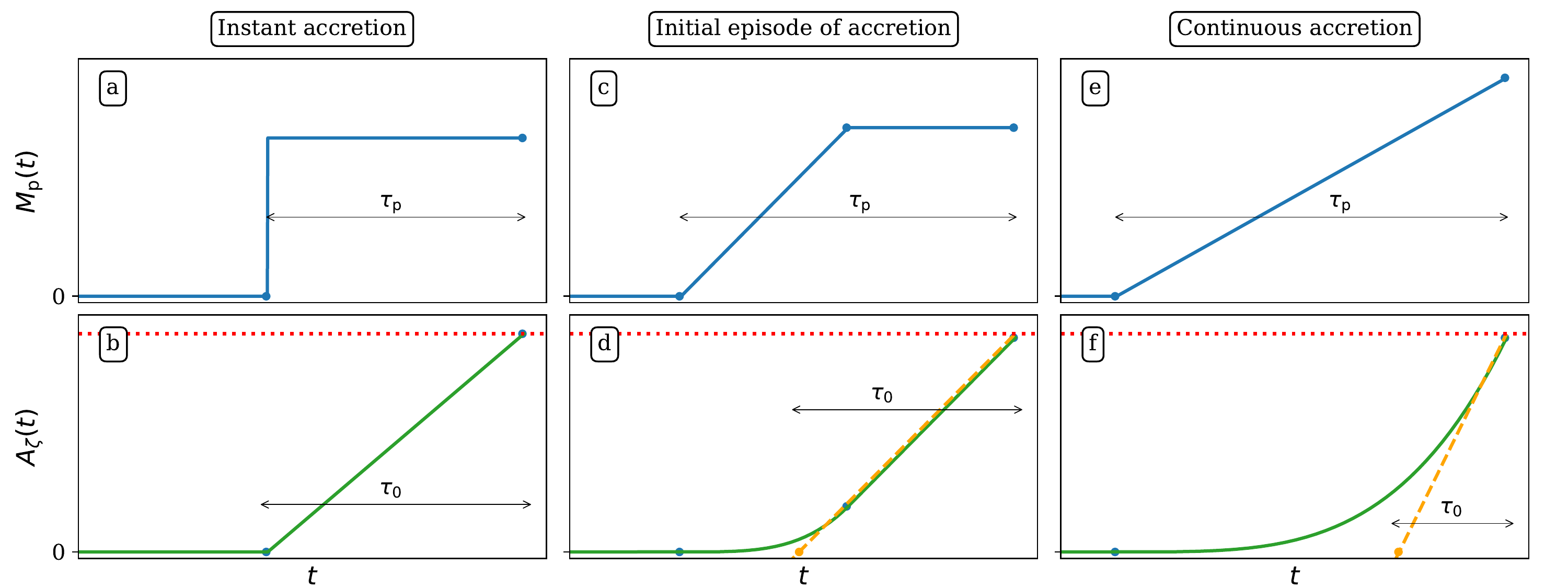}
\vspace*{0.05cm}
\caption{Illustration of the different representative planetary accretion histories: (left) very rapid (instant) initial accretion to the final mass, (centre) extended initial interval of accretion, (right) continuous accretion. Top panels illustrate $\Mp(t)$ (blue) while the bottom panels show the corresponding growth of the characteristic amplitude $A_\zeta$ (green) of the planet-driven vortensity perturbation (this calculations assumes $\de A_\zeta/\de t\propto [\Mp(t)]^{2.7}$, see text). Arrows in the top panels indicate the planetary age $\tp$ (time since the start of its accretion), while in the bottom panels they show the "time to vortex formation" $\tau_0$ calculated using equation (\ref{eq:taufit}) and assuming $\Mp$ given by its final value. The red dotted line indicates the critical value of $A_\zeta$ when the vortices are expected to appear. The key point illustrated here is that $\tau_0\le\tp$ always.
}
\label{fig:accr}
\end{figure*}

Our results (\ref{eq:taufit}) \& (\ref{eq:pars}) for $\tvrt$ have been obtained in \citetalias{Cimerman2022} for a constant $\Mp$ (not varying in time). This implicitly assumes that planet has grown to its final $\Mp$ very rapidly, having accreted its mass almost instantaneously; this accretion history is illustrated in panel (a) of Fig. \ref{fig:accr}. In panel (b) we also illustrate the corresponding growth of the characteristic amplitude\footnote{E.g. a maximum or minimum value of $\zeta$ as a function of radius, see \citetalias{Cimerman2022}.} $A_\zeta$ of the planet-induced vortensity perturbation $\zeta$, which is the variable that eventually determines vortex generation \citepalias{Cimerman2022}: very crudely, one may expect the RWI to set in when $A_\zeta$ reaches some threshold value (illustrated with red dotted line). The growth rate of $A_\zeta$ in panel (b) is constant since it sensitively depends on $\Mp$ and $\Mp$ is fixed in this case.

One may consider other representative histories of planetary mass evolution. For example, in Fig. \ref{fig:accr}c $\Mp$ undergoes an initial period of accretion and then stays at its final value until the RWI sets in. As another example, in panel (e) the planetary mass increases steadily and the RWI gets triggered while $\Mp$ is still growing. For these growth histories, the increase of $A_\zeta$ is no longer purely linear, see panels (d) and (f), and using the final planetary mass\footnote{For simplicity we neglect the possible growth of $\Mp$ after the vortex has appeared and until the present time.} in formula (\ref{eq:taufit}) we would {\it underestimate} the true age of the planet $\tp$ (illustrated in top panels), i.e. the time since its growth has started and until the present day when the vortex has emerged. Instead, application of equation (\ref{eq:taufit}) would give us some other time $\tau_0$, which is illustrated by the orange lines (based on the growth rate of $A_\zeta$ at the time when RWI sets in) in  panels (d) \& (f). Since growth of $\zeta$ accelerates (quite steeply) for higher $\Mp$, the growth rate of $A_\zeta$ can only increase in time, so that $\tp\ge\tau_0$ always (with equality only for $\Mp(t)\,=\const$, see Fig. \ref{fig:accr}a,b). 

Very importantly, this complication does not affect the validity of our time constraint, since $\tau_0$ is given by our equation (\ref{eq:taufit}) and we just saw that $\tp\ge\tau_0$. However, in some scenarios, e.g. in the continuous accretion case shown in panels (e),(f), $\tau_0$ can be much shorter than $\tp$, making our time constraint (\ref{eq:times}) too conservative. Thus, it is desirable to find ways to somehow account for the history of accretion (provided that it is known) to improve limits on $\tp$.

One way to do this has already been discussed in \citetalias{Cimerman2022} and amounts to replacing $\tp\Mp^{-\alpha}$ with $\int_0^{\tp}\left[\Mp(t)\right]^{-\alpha}\de t$ in equation (\ref{eq:combined}); this modification allows us to account for the evolution of the vortensity (or $A_\zeta$) growth rate, which is proportional to $\Mp^{-\alpha}$, as $\Mp(t)$ increases. Thus, we generalize the combined constraint on $\tp$ and $\Mp$ in the case of an accreting planet to
\begin{align}
\int_0^{\tp}\left[\Mp(t)\right]^{-\alpha} \,\de t > A \Pp \Mth^{-\alpha} \hp^\beta.
\label{eq:combined_acc}
\end{align}
Since this constraint must reduce to the inequality (\ref{eq:combined}), we will assume all its parameters --- $\alpha$, $\beta$, $A$ --- to be still given by the equation\footnote{This assumption is only approximate since the non-trivial history of accretion may modify the radial profile of $\zeta$, which determines the RWI stability \citep{Cimerman2021}. Also, the RWI threshold itself is not entirely universal \citepalias{Cimerman2022}. But this approximation should not be too bad as the RWI onset is mainly determined by the late-time behavior of $\Mp(t)$. Finally, note that theoretical arguments suggest $\alpha=2.6$, but the numerical results are closer to $\alpha\approx 2.7 $ \citepalias{Cimerman2022}.} (\ref{eq:pars}). Then in physical units equation (\ref{eq:combined_acc}) becomes
\begin{align}
\left({\rm Myr}\right)^{-1}\int_0^{\tp}\left[\frac{\Mp(t)}{10^2M_\oplus}\right]^{2.7}\de t & \,> \, 0.11 \left(\frac{\Rp}{50\mathrm{AU}}\right)^{1.5}
\nonumber\\
& \times \left(\frac{M_\star}{M_\odot}\right)^{2.2}\left(\frac{\hp}{0.1}\right)^{7.2}. 
\label{eq:combined_val_acc}
\end{align}
For $\Mp(t)=$ const this inequality reduces to (\ref{eq:combined_val}).

We can apply this generalized criterion to the simulations of \citet{Hallam2020} who considered planetary accretion history in the form $\Mp(t)=M_\mathrm{f}\sin^2\left[(\pi/2)(t/t_\mathrm{G})\right]$ (where $t_\mathrm{G}$ is the growth time) and determined the values of the final planet mass $M_\mathrm{f}$ such that the RWI would marginally set in at $t=t_\mathrm{G}$. In our notation this means setting $t_\mathrm{G}=\tvrt$. We can use our results and determine the relation between such $t_\mathrm{G}$ and $M_\mathrm{f}$ by changing inequality to equality in equation (\ref{eq:combined_acc}) and setting $\tp=t_\mathrm{G}$. We find, using the definition of $\Mth$ and introducing $q_\mathrm{f}=M_\mathrm{f}/M_\star$,
\begin{align}
t_\mathrm{G} = A\, \kappa^{-1} \Pp \, q_\mathrm{f}^\alpha \, \hp^{\beta-3\alpha},
\label{eq:Hallam}
\end{align}
where, for a particular accretion history of \citet{Hallam2020}, $\kappa=\int_0^1[\sin (\pi x/2)]^{2\alpha}dx\approx 0.33$. 

As these authors also included the effects of viscosity, which is known to delay the onset of RWI \citep{Hammer2017}, we cannot directly compare our results for $t_\mathrm{G}$ with theirs. However, if we focus on their smallest $q_\mathrm{f}=1.5\times 10^{-4}$ (since in their setup this corresponds to the lowest value of viscosity, closer to our inviscid setup) and adopt their $\hp=0.05$ and the fit parameters (\ref{eq:pars}), we find the age of the planet to satisfy $\tp\gtrsim t_\mathrm{G}\approx 39 \Pp$. This is comfortably below $t_\mathrm{G}\approx 200\Pp$ that \citet{Hallam2020} find for the same $q_\mathrm{f}$, consistent with the viscosity-driven delay. Equally importantly, had we used the equation (\ref{eq:combined}), that assumes $\Mp=$ const, instead of (\ref{eq:combined_acc}), we would have found $\tp\gtrsim 13\Pp$ (a factor of $\kappa$ lower), far less constraining than the result that we obtained accounting for the (known) accretion history. 

Given the steep dependence of the integrand in (\ref{eq:combined_val_acc}) on $\Mp$ (reflecting $\Mp$-dependence of the $\zeta$ growth rate), we expect $A_\zeta$ to increase the most when $\Mp(t)$ is close to its final value. This is indeed what we see in Fig. \ref{fig:accr}d, in which the initial accretion episode contributes only weakly to the total increase of $A_\zeta$, despite its duration being comparable to the time interval when $\Mp$ stayed at its final value (our calculation in this plot assumed $\de A_\zeta/\de t\propto \Mp^{2.7}$ for compatibility with equation (\ref{eq:combined_val_acc}), see \citetalias{Cimerman2022}). Thus, it is the history of $\Mp$ accretion at late times that is most important for determining the age of a putative planet in an observed vortex-hosting system.


\section{Application to observed discs}
\label{sec:real}

We apply the constraints derived above to several protostellar systems observed by {\it ALMA}, for which the vortices have been invoked as a possible explanation of the observed non-axisymmetric features --- arcs, clumps, etc. It is important to remember that the features detected in continuum emission by {\it ALMA} are due to thermal emission of dust grains, while our results on the emergence of vortices apply to the gaseous component of the disc. However, it has been shown by a number of authors \citep{Barge1995,Godon1999,Fu2014} that vortices are very efficient at trapping dust, providing support to our association of the dust asymmetries with the gas vortices in PPDs. Since our limits on $\tau_\mathrm{p}$ and $\Mp$ are highly sensitive to the disk aspect ratio $\hp$, which is poorly known in most cases, we will retain the scaling with $h_{0.1}=\hp/0.1$ in our estimates.\\

\noindent {\bf HD 135344B (SAO 206462)} \\
This $M_\star=1.5M_\odot$, $\tau_\mathrm{sys}\approx 9$ Myr old \citep{Asen2021} Herbig F star harbours a transitional disc. {\it ALMA} dust continuum observations reveal an axisymmetric inner ring separated by a gap-like structure (centered around $70$ AU) from an (outer) arc that can be interpreted as a vortex at the outer gap edge \citep{vdM2016,Cazzoletti2018}. The possibility of a planetary origin of these structures is supported by the near-IR scattered light observations of a two-armed spiral \citep{Muto2012}, although a unified model explaining all these features at once is lacking. We will nevertheless assume that the gap and the outer vortex are due to the (unseen) gap-opening planet at $\Rp\approx 70$ AU, and the inner ring reflects dust trapping at the pressure maximum at the inner gap edge. These data and equations (\ref{eq:M-constr})-(\ref{eq:M-vrt}) allow us to constrain planetary mass as $\Mp\gtrsim 32 h_{0.1}^{2.7}M_\oplus$.

Direct imaging of HD 135344B with VLT/SPHERE sets an upper limit of $M_\downarrow\approx 4 M_\mathrm{J}$ on the mass of a planetary object at $\sim 10^2$AU scales \citep{Asen2021}. Unfortunately, this $M_\downarrow$ is higher than the thermal mass $\Mth=1.5h_{0.1}^{3}M_\mathrm{J}$, which makes the use of the timescale constraint (\ref{eq:times}),(\ref{eq:T_vrt}) unjustified (its blind application would give $\tau_\mathrm{vrt}\approx 500h_{0.1}^{7.2}$ yr, comparable to the orbital period at the gap location and not constraining $\tp$ effectively).  \\

\noindent {\bf HD 36112 (MWC 758)} \\
This $M_\star\approx 1.8M_\odot$,  $\tau_\mathrm{sys}\approx 9$ Myr old \citep{Asen2021} star harbours two clumps on top of the two rings separated by a gap in the outer disc \citep{Dong2018}. Neglecting the slight eccentricity of the disc and assuming the rings with clumps to correspond to the inner and outer edges of the gap carved by a planet, we will adopt $\Rp\approx 70$ AU for the planetary orbit. Then equations (\ref{eq:M-constr})-(\ref{eq:M-vrt}) allow us to set a mass constraint $\Mp\gtrsim 37 h_{0.1}^{2.7}M_\oplus$.

Analysis of the direct imaging observations of this system by \citet{Asen2021} suggests that the upper limit on the possible point source inside the assumed gap is $\sim 8M_\mathrm{J}$, significantly higher than $\Mth$, precluding us from meaningfully constraining the age of the planet.\\

\noindent {\bf HD 143006} \\
This G-type T Tauri star with $M_\star=1.8M_\odot$ and an estimated age of $\tau_\mathrm{sys}\approx 8$ Myr harbours a disc rich in substructures \citep{Perez2018}. In addition to a misaligned inner disc, it features two outer rings separated by a gap centered around 52 AU, with an arc just outside the outermost ring. Interpreting these features as produced by an unseen planet inside the gap at $\Rp=52$ AU, we get the mass constraint $\Mp\gtrsim 33 h_{0.1}^{2.7}M_\oplus$ from equations (\ref{eq:M-constr})-(\ref{eq:M-vrt}). 

NaCo/VLT direct imaging does not provide a useful constraint on the mass of a putative planet, with $M_\downarrow$ at the level of several tens of $M_\mathrm{J}$ at the outer gap location \citep{Jorq2021}. Thus, we cannot set a useful lower limit on the planetary age.\\

\noindent {\bf V1247 Ori} \\
V1247 Ori is a $\tau_\mathrm{sys}=7.5$ Myr old, $M_\star\approx 1.9M_\odot$ star \citep{Wilson2019} harbouring a pre-transitional disc. {\it ALMA} dust continuum observations \citep{Kraus2017} reveal an inner disc (or ring) separated by a gap from the outer arc, which may be interpreted as a vortex at the outer gap edge. Assuming a planet to be in the gap, at $\Rp\approx 90$ AU, one finds that the planetary mass must satisfy $\Mp\gtrsim 48 h_{0.1}^{2.7}M_\oplus$.

While we could not find explicit limits on the mass of the putative planet in V1247 Ori system from direct imaging observations, \citet{Kraus2017} found that an $\Mp=3M_\mathrm{J}$ planet can roughly match the shape of the spiral observed in scattered light using HiCIAO. Unfortunately, this $\Mp$ is again above $\Mth$, not allowing the age of the planet to be meaningfully constrained.


\section{Discussion}
\label{sec:disc}



\subsection{Combination of multiple constraints}
\label{sec:comb_constr}


Our limits on $\Mp$ and $\tp$ based on the presence of vortices next to gaps in PPDs become even more powerful when combined with additional constraints on these key parameters. In particular, young planets passively lose thermal energy that they have been endowed with at formation, resulting in their luminosity decreasing with time. As more massive planets retain more heat at formation, it takes them longer to cool. Thus, if one can observationally constrain the luminosity of a planet $L_\mathrm{p}$ to lie below a certain limit (or determine it in the case of direct detection), this would provide an additional constraint on $\Mp$ and $\tp$. 

\begin{figure}
\centering
	\includegraphics[width=0.49\textwidth]{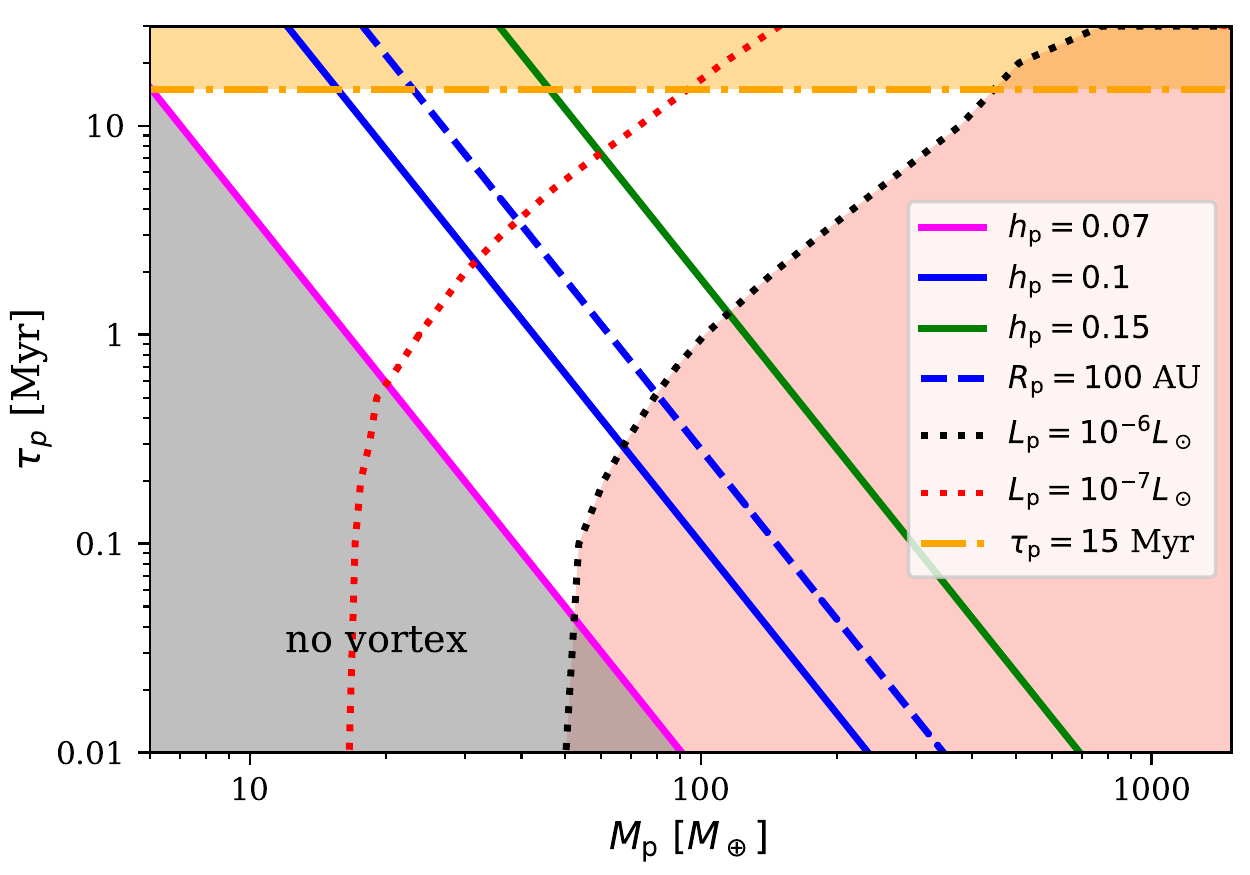}
\vspace*{-0.5cm}
\caption{Combination of the different constraints on the mass $\Mp$ and age $\tau_\mathrm{p}$ of a planet in a vortex-hosting PPD. Grey shaded region is excluded (for $\hp=0.07$) as vortices have no time to develop in this part of the parameter space, analogous to Fig. \ref{fig:age} (solid and dashed lines are the same as in that figure). Pink shaded region is excluded as it corresponds to planetary (bolometric) cooling luminosity $L_\mathrm{p}$ exceeding $L_\mathrm{p}=10^{-6}L_\odot$ (black dotted curve). We also show the curve $L_\mathrm{p}=10^{-7}L_\odot$ (red dotted curve); the $L_\mathrm{p}$ curves are based on \citet{Linder2019}. The orange shaded region above the orange dot-dashed curve excludes planetary ages above $15$ Myr. Planets satisfying all three constraints reside in the unshaded part of the parameter space. 
}
\label{fig:combined_constr}
\end{figure}

We illustrate this approach in Fig. \ref{fig:combined_constr}, where we show the vortex-based constraints from Fig. \ref{fig:age} together with the constraint $L_\mathrm{p}<10^{-6}L_\odot$ (outside the pink shaded region to the right of the black dotted curve) based on the work\footnote{We use the tracks for the bolometric luminosity $L_\mathrm{p}$ of a fixed mass planet from Fig. 6 of \citet{Linder2019}, which assume evolution with a cloud-free atmosphere of solar metallicity and use the petitCODE grid.} of \citet{Linder2019}. We also show the $L_\mathrm{p}=10^{-7}L_\odot$ curve (red dotted) which may be relevant for future direct imaging experiments. In addition, we impose a constraint $\tp < 15$ Myr (region below the orange dot-dashed line) since protoplanetary discs usually do not survive for that long (similar to the logic used in Section \ref{sec:weighting}). There are other, complementary ways of constraining planetary properties, for example gap width/depth fitting \citep{DongFung2017,Asen2021} which can provide model-dependent information on $\Mp$ for individual systems; we will not consider them here. 

A combination of the three constraints --- based on planetary luminosity, age and presence of vortices --- limits planetary $\Mp$ and $\tp$ to lie within the unshaded region. This region shrinks for hotter discs with larger $\hp$ (compare fuchsia and green solid curves), as well as for larger $\Rp$ (compare blue solid and dashed lines). Thus, the vortex-based limits are more stringent in hotter discs and for more distant planets. Also,  the allowed region would shrink even further as the upper limit on $L_\mathrm{p}$ gets lowered in the future. It should also be remembered that the luminosity-based (dotted) curves assume that the (possible ongoing) gas accretion provides insignificant contribution to $L_\mathrm{p}$. If the planetary accretion luminosity is non-negligible, this would additionally shift the dotted curves to the left, constraining $\Mp$ and $\tp$ even further. 

\subsection{Utility of the vortex-based constraint}
\label{sec:utility}


The sub-Jovian value of $M_\mathrm{vrt}$ implied by the equation (\ref{eq:M-vrt}) and our estimates in Section \ref{sec:real} is very relevant in light of the recent results \citep{Dong2017,Bae2017,Miranda2019II,Miranda2020I,Miranda2020II} showing that in a low viscosity disc a single sub-$\Mth$ planet can give rise to a series of several prominent gaps and rings in the radial dust distribution. For example, for the AS 209 system ($M_\star=0.83M_\odot$, $\tau_\mathrm{sys}\approx 1$ Myr, \citealt{Andrews2018}) imaged with {\it ALMA} \citet{Zhang2018} have shown that a single planet with $\Mp$ as low as $25M_\oplus$ orbiting within the outer (primary) gap at $\Rp\approx 100$AU can be responsible for creating all five gaps observed in this disc. This possibility makes the typical values of $\Mp$ implied by the constraint (\ref{eq:M-constr})-(\ref{eq:M-vrt}) very interesting for understanding the architecture of the underlying planetary system.

Given the upper limits on $\Mp$ based on direct imaging in several systems covered in Section \ref{sec:real}, we found our age constraint (\ref{eq:times}) \& (\ref{eq:T_vrt}) to be not very useful at present. However, things will improve as $M_\downarrow$ decreases in the future. Once $M_\downarrow$ is below $\Mth$, our constraint (\ref{eq:times}) \& (\ref{eq:T_vrt}) becomes valid and may provide useful information on the planetary age. The decrease of $M_\downarrow$ may not necessarily come from improved direct imaging capabilities. In particular, one may use the technique of multiple gap fitting used by \citet{Zhang2018} for AS 209 to get a much better measurement of $\Mp$ or $M_\downarrow$. 

Just for illustration, let us imagine that AS 209 did possess a vortex at the edge of its outermost gap (just outside $\Rp=100$AU). Then using $\Mp\approx 25M_\oplus$ (based on \citealt{Zhang2018}) equation (\ref{eq:T_vrt}) would predict $\tau_\mathrm{p}\gtrsim 8h_{0.1}^{7.2}$ Myr. This $\tp$ is much longer (for $\hp=0.1$) than the age of the system $\tau_\mathrm{sys}\approx 1$ Myr, and could have implied that either the planetary mass is underestimated (by a factor of $\sim 2$), or that the stellar age is underestimated (by almost an order of magnitude), or that the disc is somewhat colder --- using $\hp=0.075$ (consistent with \citealt{Pane2022}) in (\ref{eq:T_vrt}) would reconcile $\tp$ with its estimated $\tau_\mathrm{sys}$.

The latter possibility represents a simple way to resolve the age discrepancy for this imaginary AS 209-like system. It also highlights the importance of good knowledge of the thermal state of the disc near the planet, which sets $\hp$. Indeed, $\tvrt$ depends very sensitively on $\hp$ and a mis-estimate of $\hp$ by a factor of 2 would result in a factor of $\approx 150$ error in the determination of $\tvrt$ and the planetary age. The situation is somewhat improved for the mass constraint (\ref{eq:M-constr})-(\ref{eq:M-vrt}), in which variation of $\hp$ by a factor of 2 results in $M_\mathrm{vrt}$ changing by a factor of $\approx 6.5$. In any case, good understanding of disc thermodynamics is clearly needed when applying the age constraint (\ref{eq:times}) \& (\ref{eq:T_vrt}). Recent {\it ALMA} measurements of emission heights of different molecular lines in PPDs \citep{Law2021,Law2022,Pane2022} provide a (model-dependent) way to determine disc aspect ratio at different radii, generally finding values in the range $\hp\sim (0.07-0.1)$ for $\Rp\sim (50-100)$ AU.

On the other hand, our constraints (\ref{eq:combined_val}),(\ref{eq:M-vrt}) \& (\ref{eq:T_vrt}) should be rather insensitive to the radial profile of the disc surface density near the planet. Indeed, \citetalias{Cimerman2022} showed that the parameters of the fit (\ref{eq:taufit}),(\ref{eq:pars}) show little variation when changing the slope of the surface density profile near the planet. Also, the dependence of the vortex-based constraints on $\Rp$ and $M_\star$ is not as steep as for $\hp$, and the characteristic accuracy with which these parameters can be measured is $(10-20)\%$ or better.


\subsection{Additional processes and further extensions}
\label{sec:extensions}


Since the constraints (\ref{eq:times})-(\ref{eq:M-constr}) are {\it lower} limits on $\tau_\mathrm{p}$ and $\Mp$, respectively, they do not change if the vortices we observe in discs now are not the first generation vortices. It is possible that the vortices that formed early on have then dissolved and what we are seeing now are the second (or multiple) generation vortices \citep{Hammer2021}. Nevertheless, even in this case the condition $\tau_\mathrm{p}>\tvrt$ would still need to be fulfilled, definitely for the first generation of vortices, as well as for the following generations, confirming the validity of the constraints (\ref{eq:times})-(\ref{eq:M-constr}). 

Similarly, dust trapped in vortices can maintain observable non-axisymmetric distribution even after the vortices in the gaseous component dissolve \citep{Fu2014}. Thus, when we see an asymmetry in dust continuum observations, the original vortex that has led to it may have already been gone. However, this would again not invalidate the constraints obtained in Sections \ref{sec:weighting} \& \ref{sec:dating}. 

The fit (\ref{eq:taufit}),(\ref{eq:pars}) for $\tvrt$ was derived by \citetalias{Cimerman2022} for discs which are inviscid or have low viscosity, an assumption which is consistent with observations of many systems (see Section \ref{sec:obs}). We can roughly estimate the upper limit on the viscosity $\nu$ below which the inviscid assumption should be valid by demanding the timescale on which the vortensity structures produced by the planet get viscously diffused away to be longer that the age of the system $\tau_\mathrm{sys}$. For the characteristic radial scale of the vortensity structures $L\sim \Hp(\Mp/\Mth)^{-0.4}$ (\citealt{Dong2011II}; \citetalias{Cimerman2022}) this timescale is $\sim L^2/\nu\sim \Pp\alpha^{-1}(\Mp/\Mth)^{-0.8}$, where we adopted the $\alpha$-ansatz for the viscosity $\nu=\alpha\Omega_\mathrm{p}\Hp^2$ (and $\Omega_\mathrm{p}=2\pi\Pp^{-1}$). For this to exceed $\tau_\mathrm{sys}$ for a sub-thermal mas planet we require that roughly $\alpha\lesssim \Pp/\tau_\mathrm{sys}\sim 10^{-4}$, given the long orbital periods at $\Rp=50-100$ AU. A more refined estimate of the critical $\alpha$ can be found in \citetalias{Cimerman2022}.

However, even if the disc were sufficiently viscous (i.e. for $\alpha\gtrsim 10^{-4}$), the RWI development would get only delayed \citep{Hallam2020} or the instability may be suppressed altogether, see \citet{Hammer2017}, \citetalias{Cimerman2022}. Because of that, our inviscid estimate for $\tvrt$ continues to provide a lower limit on $\tp$ in the presence of a vortex, i.e. the equation (\ref{eq:times}) and all other constraints remain valid (see Section \ref{sec:growth} for application of this logic).

On the other hand, some other effects may {\it accelerate} vortex production compared to the results of \citetalias{Cimerman2022}. For example, this could happen as a result of baroclinicity of the disc near the planet since the RWI is sensitive to entropy gradients \citep{Lovelace1999}. Under certain circumstances dust feedback can also promote vortex production \citep{Lin2017}. These processes, if they are important, may somewhat weaken our constraints on $\Mp$ and $\tp$.

We demonstrated in Section \ref{sec:growth} how our constraints can be modified to account for the evolution of planetary mass $\Mp$. Other relevant parameters might change as well, for example $\Rp$ can vary as a result of planet migration, or $\hp$ can change as the disc evolves in time. \citetalias{Cimerman2022} outlined ways in which one can account for these processes to derive a new estimate for $\tvrt$ instead of (\ref{eq:taufit}),(\ref{eq:pars}), thus providing a pathway to modifying our constraints on $\Mp$ and $\tp$.

Of the four systems considered in Section \ref{sec:real}, three show vortex-like non-axisymmetries only at the outer edge of the putative planetary gap, and only one, MWC 758, has them on both sides of the gap. This is somewhat surprising, since the simulations of \citetalias{Cimerman2022} not only show the emergence of vortices on both sides of the gap, but also demonstrate that the time interval separating their production by RWI is typically smaller than $\tvrt$ (see Table 1 in that work). Thus, one would expect to see vortices on both sides of the gap more often. It is not clear why this expectation fails. It could be that the dust concentration is more efficient in the outer vortices\footnote{Outer vortices should form first in inviscid discs with radially decreasing surface density \citepalias{Cimerman2022}.} or that it tends to survive there considerably longer than in the inner ones. Or that some physical processes neglected in our study suppress the formation of the inner vortices. Expanding the sample of observed discs with vortex-like asymmetries would help in resolving this issue in the future.


\section{Summary}
\label{sec:summ}


In this work we used the results of \citetalias{Cimerman2022} on the time it takes visible gas vortices to appear next to a gap carved by a low-mass planet in a low-viscosity PPD to set constraints on the masses $\Mp$ and ages $\tp$ of planets in PPDs with observed vortex-like structures. We found that the presence of a vortex sets a lower limit on a particular combination of $\Mp$ and $\tp$, with separate constraints on these variables possible if some additional information (such as the system age $\tau_\mathrm{sys}$ or the upper limit on the planetary mass $M_\downarrow$) is available. These considerations allowed us to constrain the masses of putative planets in several vortex-bearing PPDs to be above several tens of $M_\oplus$. The limits on the planetary age are not very constraining at the moment, but they will improve as future observations lower $M_\downarrow$. Our constraints can be extended to account for the non-trivial history of planetary mass accretion, and we provide a recipe for doing that in Section \ref{sec:growth}. Finally, we showed the robustness of our constraints in light of additional complications (e.g. non-zero disc viscosity, multiple generation of vortices, etc.) and demonstrated their useful synergy with other types of constraints on $\Mp$ and $\tp$, e.g. based on the upper limits on the planetary cooling luminosity coming from direct imaging observations.


\section*{Acknowledgements}

\textit{Software:} Matplotlib \citep{Matplotlib}. Authors are grateful to Ewine van Dishoeck for illuminating discussions and to an anonymous referee for useful suggestions. R.R.R. acknowledges financial support through the Science and Technology Facilities Council (STFC) grant ST/T00049X/1 and Ambrose Monell Foundation.  N.P.C. is funded by a STFC and Isaac Newton studentship.

\section*{Data Availability}
The data underlying this article will be shared on reasonable request to the corresponding author.




\bibliographystyle{mnras}
\bibliography{example} 





\bsp	
\label{lastpage}
\end{document}